\documentclass[a4paper,12pt]{paper}
\pdfoutput=1
\usepackage[latin1]{inputenc}
\usepackage[T1]{fontenc}
\usepackage[english]{babel}
\usepackage{amssymb}
\usepackage{amsmath}
\usepackage{trfsigns}
\usepackage{graphicx}
\usepackage{hyperref}
\usepackage[numbers,sort&compress]{natbib}

\DeclareGraphicsExtensions{.pdf,.png,.jpg,.eps}
\DeclareMathOperator{\uimm}{\mathrm{i}}
\title{Predictions of a model of weak scale from dynamical breaking of scale invariance}
\author{\emph{Giulio Maria Pelaggi}\\Dipartimento di Fisica dell'Universit\`a di Pisa and INFN, Pisa, Italy\\\small{g.pelaggi@for.unipi.it}}
\date{}

\begin{document}
\pagestyle{plain}
\maketitle
\abstract
We consider a model where the weak and the Dark Matter scale arise at one loop from the Coleman-Weinberg mechanism, where the spontaneous symmetry breaking is driven by the Higgs and a new scalar boson. The latter is a doublet under the new gauge group $\text{SU(2)}_X$ of the lagrangian. The new gauge vector is a good candidate to represent the Dark Matter particle. We perform a precise computation of the model predictions for the production cross section of the new scalar and for the direct-detection cross section of the DM particle, as a function of the only free parameter of the theory. The model can be tested directly in the next run of LHC and in the experiments LZ and XENON1T, because it provides particles in a slightly higher range of energies than the actual experimental limits.
\tableofcontents

\section{Introduction}
One year ago, the first LHC run ended. Its most important result is the discovery of a particle, with a mass of about 126 GeV, that is fully compatible with the Higgs boson of the Standard Model (SM) \cite{atlas, cms}. This data confirms that the predictions on the SM spectrum were true, but the success of the SM hides another result: the complete absence of new physics.

Though the SM explains a wide range of physical phenomena, it has still some unsatisfactory aspects.

One of them is known with the name of ``hierarchy problem'' \cite{thoft}. Why the dimensional parameters of the SM, that is the Higgs mass, has to get the value that we can measure experimentally? The SM doesn't explain the link between the Higgs mass and other fundamental energy scales, e.g. the Planck mass. To describe our world, we need to ``fine-tune'' the parameters, that is, we have to set precisely their bare value to reproduce the experimental value.

A guideline for the physics beyond the SM may be naturalness \cite{thoft}. Fundamentally we introduce some new physics, such that the divergent parts of the correction to the Higgs mass given by the SM are canceled by the new particles and interactions, while the remaining corrections are smaller than $M_h^2$. In other words, naturalness suggests the existence of new physics at a certain scale $\Lambda_\text{nat}$. The corrections due to the SM are $\delta M_h^2 \sim \Lambda_\text{nat}^2$, so if we want small corrections with respect to $M_h^2$, the new physics has to be at low energies. Therefore, we hope that the new physics could explain the origin of the unpredicted values of the parameters of the Standard Model.

Looking at the data of the last period, i.e.~the absence of these particles around the weak scale, the scale of the new physics had to move towards greater energy values. Maybe naturalness is a wrong way, but we think that some aspects of it can be recovered.

The model we will consider in this article is not UV complete, so the new physics added to the Standard Model doesn't cancel automatically the divergences to the Higgs mass.
The ``finite naturalness'' is a way to explain how to deal with these divergences. We think that the unknown physical cut-off behaves like in dimensional regularization computations: the divergent parts of the correction are unphysical, so we can neglect them. The reliability of finite naturalness for the SM has been studied in \cite{fari}.

The fundamental idea is to start from a model with a lagrangian that doesn't have any mass terms for scalar particles: the Coleman-Weinberg mechanism provides us a method to explain a non-zero value of the masses, also if the mass term is null, considering the radiative corrections of the theory. In fact, Coleman and Weinberg explain that the Spontaneous Symmetry Breaking (SSB) is not necessarily driven by a negative mass term for the scalar particle, but it can arise because of high-order processes involving virtual particles \cite{coleman}. In fact, if we consider the one-loop effective potential of our model, new minima arise and SSB occurs.

In this way we get rid of the presence of dimensional parameters that make the lagrangian not scale invariant, and, following the idea of the finite naturalness, we don't get quadratic divergences.
Some models with this peculiarity has already been studied, for example in \cite{tipocoleman1, tipocoleman2, tipocoleman3, tipocoleman4, tipocoleman5, tipocoleman6, tipocoleman7, tipocoleman8, tipocoleman9, tipocoleman10, tipocoleman12, estonia1, estonia2, estonia3}.

The SM doesn't provide a description of the Dark Matter. Today we can ``see'' it only through gravitational interactions. If we want to describe it as a particle, none of the particles we already know are good candidates.

The model that we are going to study has been proposed in \cite{hamb, stru, carone}: the stability of the DM particle is not given by an  \textit{ad hoc} symmetry, but only because the gauge symmetry of the lagrangian and because of the particle content of the theory. The same happens in the SM, where the photon, the electron, the proton and the lightest neutrino are stable. In this model, the DM particle is a multiplet of vector particles.
Following this idea, we want to introduce a new hidden sector of the lagrangian, that is connected to the SM only through a scalar quartic interaction with the Higgs boson (Higgs portal). For this purpose, we have to introduce a new scalar particle.

In the following sections, we compute more precisely if this model can be confirmed by the experimental data, maybe in the next phase of the work of LHC. We will analyze the general properties of the model in Section \ref{themodel}, in Section \ref{completecomputation} we show all the computations done in the most general case, while in Section \ref{results} we show the results and we discuss them.

\section{The model}\label{themodel}
In this section we will briefly study the properties of this model. In particular we will explain why should we study the one-loop effective potential of the theory.

\subsection{Lagrangian and particle content}
We choose SU(2)$_\text{X}$ as the new gauge group of the lagrangian, so the entire model is symmetrical under U(1)$_\text{Y}\times$SU(2)$_\text{L}\times$SU(3)$_\text{c}\times$SU(2)$_\text{X}$. The particle content is given by the SM particle content; plus we introduce doublet $S$ of the group SU(2)$_\text{X}$, that is a Lorentz scalar and a singlet under the SM symmetry group. We call $X_\mu$ the SU(2)$_\text{X}$ vectors. These particles are, according to the model, the ones that constitute the Dark Matter. $X_\mu$ bosons can be described as $X_\mu=X_\mu^a T^a$, where $T^a$s are the generators of the new symmetry group, and they have a kinetic lagrangian term $-\frac{1}{4}F_{\mu\nu}^XF^{\mu\nu}_X$, where $F_{\mu\nu}^X=[D_\mu,D_\nu]$. The kinetic term of the new scalar field is $|D_\mu S|^2$, where $D_\mu=\partial_\mu+\uimm g_X X_\mu$ and $g_X$ is the new coupling constant of the SU(2)$_\text{X}$ gauge group.
For more details see~\cite{stru}

\subsection{Tree-level potential}
The tree-level scalar potential $V_0$ is
\begin{equation}V_0=\lambda_H |H^\dagger H|^2-\lambda_{HS}|H^\dagger H||S^\dagger S|+\lambda_S |S^\dagger S|^2.\end{equation} 
We observe that there is not a mass term for the Higgs field nor for the new scalar boson: as we said before, they will get their mass through the Coleman-Weinberg mechanism, considering one loop contributes to the theory.
The SSB down to U(1)$_\text{em}\times$SU(3)$_\text{c}$ occurs, and so the degrees of freedom represented by the six Goldstone bosons of the theory are absorbed into the longitudinal polarizations of all the gauge bosons. We can expand the scalar field in components as
\begin{equation}H(x)=\frac{1}{\sqrt{2}}\begin{pmatrix}0\\v+h(x)\end{pmatrix},\qquad S(x)=\frac{1}{\sqrt{2}}\begin{pmatrix}0\\w+s(x)\end{pmatrix}.\end{equation}
SU(2)$_\text{X}$ is broken by the VEV $w$ of the doublet $S$, so every $X_\mu$ boson gets the same mass $M_X=g_X w/2$ from the interaction with the $S$ field.

The new lagrangian parameters introduced are $\lambda_S$, $\lambda_H$, $\lambda_{HS}$ and $g_X$.

\section{Precise computation}\label{completecomputation}
We aim to find the values of these parameters and of $v$ and $w$ in terms of some known experimental data, so we compute some appropriate observables like the Higgs mass, the annihilation and semiannihilation cross sections of the DM, the muon decay amplitude.

\subsection{One-loop potential}
To find a minimum point different from the origin we have to consider the one-loop contributions computing the one-loop potential. The result for this theory is
\begin{equation}V^{\text{1loop}}=V_0+V_1\end{equation}
\begin{align*}V_1=\frac{1}{64\pi^2}\Biggl[3 f_{5/6}(m_Z^2)-f_{3/2}(\xi_Z m_Z^2)+6f_{5/6}(m_W^2)-2f_{3/2}(\xi_W m_W^2)+\biggr.\\\left.+9f_{5/6}(m_X^2)-3f_{3/2}(\xi_X m_X^2)-12f_{3/2}(m_t^2)+\sum_i{f_{3/2}(m_i)}\right].\end{align*}
The expression for the $f$ function is
\begin{equation}f_c(x)=x^2\left(\frac{1}{\epsilon}+\ln{\frac{x}{\mu^2}}-c\right),\end{equation}
where $\mu$ is the energy scale at which we are renormalizing the theory.
We consider Z, W and X loops (with longitudinal polarization for every vector), quark top loops (we suppose that this quark is the only fermion that gives a contribution), scalar particles and Faddeev-Popov ghosts loops. The sum is over all the scalar particles of the theory, that is the six Goldstone bosons and the two scalars $h$ and $s$. In this expression, $\xi_Z$, $\xi_W$ and $\xi_X$ are the parameters that determine the gauge fixing for the Z, W and X sectors, respectively. We will choose the Landau gauge, so we will take  $\xi_Z=\xi_W=\xi_X=0$. The three expressions for the masses of the Goldstone bosons related to the $H$ field are $m_{1,2}=v^2\lambda_H-w^2\lambda_{HS}/2$ and $m_3=v^2\lambda_H-w^2\lambda_{HS}/2$, while for the three Goldston bosons of the $S$ field we have $m_{4,5,6}=w^2\lambda_S-v^2\lambda_{HS}/2$. Regarding the masses of the two physical scalars, we observe that the tree-level mass matrix is not diagonal:
\begin{equation}M_0=\begin{pmatrix}3 v^2 \lambda_H - w^2\lambda_{HS}/2 && -vw\lambda_{HS}\\-vw\lambda_{HS} && 3 w^2 \lambda_S - v^2\lambda_{HS}/2 \end{pmatrix}.\end{equation}
Since we want to describe scalar fields using the eigenstates of this matrix, the eigenvalues are their masses:
\begin{align*} \tilde{m}_{1,2}= \frac{1}{4}&\Bigl[v^2(6\lambda_H-\lambda_{HS})-w^2(\lambda_{HS}-6\lambda_S)\biggr.\\ &\pm\left(-2v^2w^2\left(\lambda_{HS}(6\lambda_H-7\lambda_{HS})+6\lambda_S(6\lambda_{H}+\lambda_{HS})\right)\right.\\ &\quad\left.\left. +v^4(6\lambda_H+\lambda_{HS})^2+w^4(\lambda_{HS}+6\lambda_S)^2\right)^{1/2} \right].
\end{align*}
The interaction eigenstates don't coincide with mass eigenstates. A mixing angle that correlates the two basis will be introduced.

\subsection{Minimum equations}\label{minimumequations}

We can choose freely the energy scale of the renormalized theory. Depending of the values of the running parameter of the model, we can have a potential with a minimum in the origin if $4\lambda_H\lambda_S-\lambda_{HS}^2>0$, or a saddle point in the other case. To simplify calculations, we choose the critical scale where $4\lambda_H\lambda_S-\lambda_{HS}^2=0$. In this situation, the tree-level potential has minima on two straight lines passing for the origin:
\begin{equation}\frac{v}{w}=\left(\frac{\lambda_H}{\lambda_S}\right)^{1/4}.\end{equation}
As a consequence, one of the two scalar masses is null.
This choice is possibile because, if we study the running of the constants as a function of the energy~\cite{stru}, we can see that there is an energy $\mu_\ast$ where this condition is satisfied. At this point, we can replace $\lambda_S$ by $\mu_\ast$, so the parameters of the theory become $\lambda_H, \lambda_{HS}$ and the critical scale energy $\mu_\ast$.

Now we switch to the effective potential. We impose that the first derivatives of the potential with respect to the fields cancel:
\begin{align*}\frac{\partial V}{\partial v}=v(\lambda_H v^2 - w^2\lambda_{HS}/2)+ T_h=0,\\\frac{\partial V}{\partial w}=w(\lambda_S w^2 - v^2\lambda_{HS}/2)+ T_s=0,\end{align*}
where $T_h$ and $T_s$ represent the one loop tadpoles related to the two scalars.

\subsection{Scalar masses}\label{massaHiggs}

We have seen that the tree-level mass matrix is not diagonal and we call its eigenstates $h_1$ and $h_2$. At one-loop approximation, the pole masses of the two physical scalars are the values of $p$ for which
\begin{equation}\det\begin{pmatrix} p^2-\tilde{m}^2_{1}-\Pi_{11}(p^2) && -\Pi_{12}(p^2)\\ -\Pi_{12}(p^2) && p^2-\tilde{m}^2_{2}-\Pi_{22}(p^2)\end{pmatrix}=0,\end{equation}
where $\tilde{m}_{1,2}$ represent the tree-level masses of the scalars, while $\Pi_{ij}(p^2)$ represents the one-loop corrections to the propagator at the energy $p^2$.
If we want to consider only the one-loop approximation, off-diagonal terms are not important, and can be neglected, so the pole masses are
\begin{equation}M_1^2=\tilde{m}_1^2+\Pi_{11}(\tilde{m}_1^2),\qquad M_2^2=\tilde{m}_2^2+\Pi_{22}(\tilde{m}_2^2).\end{equation}

In the critical condition we have chosen, one of the tree-level masses of the two scalars cancels, so the correction to it wouldn't be a small perturbation anymore, but it would constitute the entire value of the observable. Because of this, we compute the one-loop correction of the masses in two subsequent steps. We split $\Pi(p^2)$ in two parts:
\begin{equation}\Pi(p^2)=\Pi(0)+\Delta\Pi(p^2).\end{equation}
For both diagonal elements and off-diagonal ones, we can obtain $\Pi(0)$ computing the second derivatives of $V$ with respect to the fields. $\Delta\Pi(p^2)$ is the contribution due to the wavefunction renormalization. Given that only $\Pi(0)$ is proportional to the masses of the heaviest particles, and so $\Delta\Pi(p^2)\ll \Pi(0)$, we can have a good approximation for the scalar masses computing again the eigenvalues, but this time we neglect the wavefunction correction. We call this eigenvalues $m_1$ and $m_2$.
As we said before, the off-diagonal terms of these corrections are not important, so the final expressions for the masses of the scalars are:
\begin{equation}M_1^2=m_1^2+\Delta\Pi_{11}(m_1^2),\qquad M_2^2=m_2^2+\Delta\Pi_{22}(m_2^2).\end{equation}
We observe that the one-loop potential doesn't take into account the renormalization of the wavefunction. To compute this correction we have to start from the one-loop correction to the propagators of $h_1$ and $h_2$.
More precisely, we can write $\Delta\Pi(p^2)=\Pi(p^2)-\Pi(0)$, so we compute the one-loop contributions to the two-points Green function of each mass eigenstate for a generic $p^2$ and for $p=0$ and than we do the subtraction.

We introduce the mixing angle $\alpha$, that is the rotation angle needed to diagonalize the one-loop mass matrix. It is defined by the relations
\begin{equation}h_1 = h\cos{\alpha} + s\sin{\alpha} \text{\quad and\quad} h_2 = s\cos{\alpha} - h\sin{\alpha}.\end{equation}

The expression for the Higgs one-loop propagator is

\begin{align*}
&\Pi(p^2)=2\left(\frac{3 g_2^2 A_0\left(M_W^2\right)}{64 \pi ^2}-\frac{g_2^2 M_W^2}{32 \pi ^2}\right) \cos ^2\alpha\addtocounter{equation}{1}\tag{\theequation}\\
&+\left(\frac{3 g_2^2 M_Z^2 A_0\left(M_Z^2\right)}{64 M_W^2 \pi ^2}-\frac{g_2^2 M_Z^4}{32 M_W^2 \pi ^2}\right) \cos ^2\alpha+3 \left(\frac{3 g_X^2 A_0\left(M_X^2\right)}{64 \pi ^2}-\frac{g_X^2 M_X^2}{32 \pi ^2}\right) \sin ^2\alpha\\
&-\left(\frac{3 g_2^2 M_t^2 A_0\left(M_t^2\right)}{16 M_W^2 \pi ^2}+\frac{3  g_2^2\left(4 M_t^4-M_t^2 p^2\right) B_0\left(p^2,M_t^2,M_t^2\right)}{32 M_W^2 \pi ^2}\right) \cos ^2\alpha\\
&+2 \left(-\frac{\left(M_W^2+p^2\right) A_0\left(M_W^2\right) g_2^2}{64 M_W^2 \pi ^2}\right.\\
&\left.\qquad+\frac{ g_2^2\left(M_W^4-2 p^2 M_W^2+p^4\right) B_0\left(p^2,M_W^2,0\right)}{64 M_W^2 \pi ^2}-\frac{ g_2^2p^4 B_0(p^2,0,0)}{64 M_W^2 \pi ^2}\right) \cos ^2\alpha\\
&+\left(\frac{A_0\left(M_W^2\right) g_2^2}{32 \pi ^2}+\frac{ g_2^2p^4 B_0(p^2,0,0)}{64 M_W^2 \pi ^2}+\frac{ g_2^2\left(12 M_W^4-4 p^2 M_W^2+p^4\right) B_0\left(p^2,M_W^2,M_W^2\right)}{64 M_W^2 \pi ^2}\right.\displaybreak[0]\\
&\left.\qquad-\frac{M_W^2 g_2^2}{8 \pi ^2}-\frac{g_2^2\left(M_W^4-2 p^2 M_W^2+p^4\right) B_0\left(p^2,M_W^2,0\right) }{32 M_W^2 \pi ^2}\right) \cos ^2\alpha\\
&+\left(-\frac{ g_2^2\left(M_Z^2+p^2\right) A_0\left(M_Z^2\right)}{64 M_W^2 \pi ^2}\right.\\
&\left.\qquad+\frac{g_2^2\left(M_Z^4-2 p^2 M_Z^2+p^4\right) B_0\left(p^2,M_Z^2,0\right) }{64 M_W^2 \pi ^2}-\frac{ g_2^2p^4 B_0(p^2,0,0)}{64 M_W^2 \pi ^2}\right) \cos ^2\alpha\\
&+\left(\frac{ A_0\left(M_Z^2\right)g_2^2 M_Z^2}{64 M_W^2 \pi ^2}+\frac{g_2^2 p^4 B_0(p^2,0,0)}{128 M_W^2 \pi ^2}+\frac{g_2^2 \left(12 M_Z^4-4 p^2 M_Z^2+p^4\right) B_0\left(p^2,M_Z^2,M_Z^2\right)}{128 M_W^2 \pi ^2}\right.\\
&\left.\qquad-\frac{g_2^2 M_Z^4}{16 M_W^2 \pi ^2}-\frac{g_2^2 \left(M_Z^4-2 p^2 M_Z^2+p^4\right) B_0\left(p^2,M_Z^2,0\right)}{64 M_W^2 \pi ^2}\right) \cos ^2\alpha\\
&+3 \left(-\frac{ g_X^2\left(M_X^2+p^2\right) A_0\left(M_X^2\right)}{64 M_X^2 \pi ^2}\right.\\
&\left.\qquad+\frac{g_X^2\left(M_X^4-2 p^2 M_X^2+p^4\right) B_0\left(p^2,M_X^2,0\right) }{64 M_X^2 \pi ^2}-\frac{ g_X^2p^4 B_0(p^2,0,0)}{64 M_X^2 \pi ^2}\right) \sin ^2\alpha\displaybreak[0]\\
&+3 \left(\frac{A_0\left(M_X^2\right) g_X^2}{64 \pi ^2}+\frac{ g_X^2p^4 B_0(p^2,0,0)}{128 M_X^2 \pi ^2}+\frac{g_X^2\left(12 M_X^4-4 p^2 M_X^2+p^4\right) B_0\left(p^2,M_X^2,M_X^2\right) }{128 M_X^2 \pi ^2}\right.\\
&\left.\qquad-\frac{M_X^2 g_X^2}{16 \pi ^2}-\frac{ g_X^2\left(M_X^4-2 p^2 M_X^2+p^4\right) B_0\left(p^2,M_X^2,0\right)}{64 M_X^2 \pi ^2}\right) \sin ^2\alpha\displaybreak[0]\\
&+\frac{3 B_0(p^2,0,0)}{32 \pi ^2} (2 w \lambda_S \sin \alpha-v \lambda_{HS} \cos \alpha)^2+\frac{3 B_0(p^2,0,0)}{32 \pi ^2} (2 v \lambda_H \cos \alpha-w \lambda_{HS} \sin \alpha)^2\displaybreak[0]\\
&+\frac{B_0\left(p^2,M_{h_1}^2,M_{h_2}^2\right)}{32 \pi ^2} \left(v \lambda_{HS} \cos ^3\alpha-2 w \lambda_{HS} \sin \alpha \cos ^2\alpha-6 w \lambda_S \sin \alpha \cos ^2\alpha\right.\\
&\left.\qquad\qquad\qquad\qquad-6 v \lambda_H \sin ^2\alpha \cos \alpha-2 v \lambda_{HS} \sin ^2\alpha \cos \alpha+w \lambda_{HS} \sin ^3\alpha\right)^2\\
&+\frac{B_0\left(p^2,M_{h_2}^2,M_{h_2}^2\right)}{32 \pi ^2} \left(v \lambda_{HS} \cos ^3\alpha+6 v \lambda_H \sin \alpha \cos ^2\alpha+2 v \lambda_{HS} \sin \alpha \cos ^2\alpha\right.\\
&\left.\qquad\qquad\qquad\qquad-2 w \lambda_{HS} \sin ^2\alpha \cos \alpha-6 w \lambda_S \sin ^2\alpha \cos \alpha-w \lambda_{HS} \sin ^3\alpha\right)^2\displaybreak[0]\\
&+\frac{9 B_0\left(p^2,M_{h_1}^2,M_{h_1}^2\right)}{32 \pi ^2} \left(2 v \lambda_H \cos ^3\alpha+w \lambda_S \sin ^3\alpha\right.\\
&\left.\qquad\qquad\qquad\qquad-\lambda_{HS} \left(w \sin \alpha \cos ^2\alpha+v \sin ^2\alpha \cos \alpha\right)\right)^2\\
&+\frac{3 A_0\left(M_{h_1}^2\right) \left(\lambda_H \cos ^4\alpha-\lambda_{HS} \sin ^2\alpha \cos ^2\alpha+\lambda_S \sin ^4\alpha\right)}{16 \pi ^2}\\
&+\frac{A_0\left(M_{h_2}^2\right) \left((6 \lambda_H+4 \lambda_{HS}+6 \lambda_S) \cos ^2\alpha \sin ^2\alpha-\lambda_{HS} \left(\cos ^4\alpha+\sin ^4\alpha\right)\right)}{32 \pi ^2}.
\end{align*}

In this formula $A_0$ and $B_0$ are the Passarino-Veltman functions:
\begin{equation}A_0(m^2)=\frac{1}{\uimm\pi^{D/2}}\int{\text{d}q^D\frac{1}{q^2-m^2+\uimm\epsilon}}\end{equation}
\begin{equation}B_0(p^2,m_1^2,m_2^2)=\frac{1}{\uimm\pi^{D/2}}\int{\text{d}q^D\frac{1}{(q^2-m^2+\uimm\epsilon)((q+p)^2-m_2^2+\uimm\epsilon)}}.\end{equation}

\subsection{Dark Matter abundance}\label{DMabundance}
\begin{figure}[bht]
\centering
\includegraphics[scale=0.4]{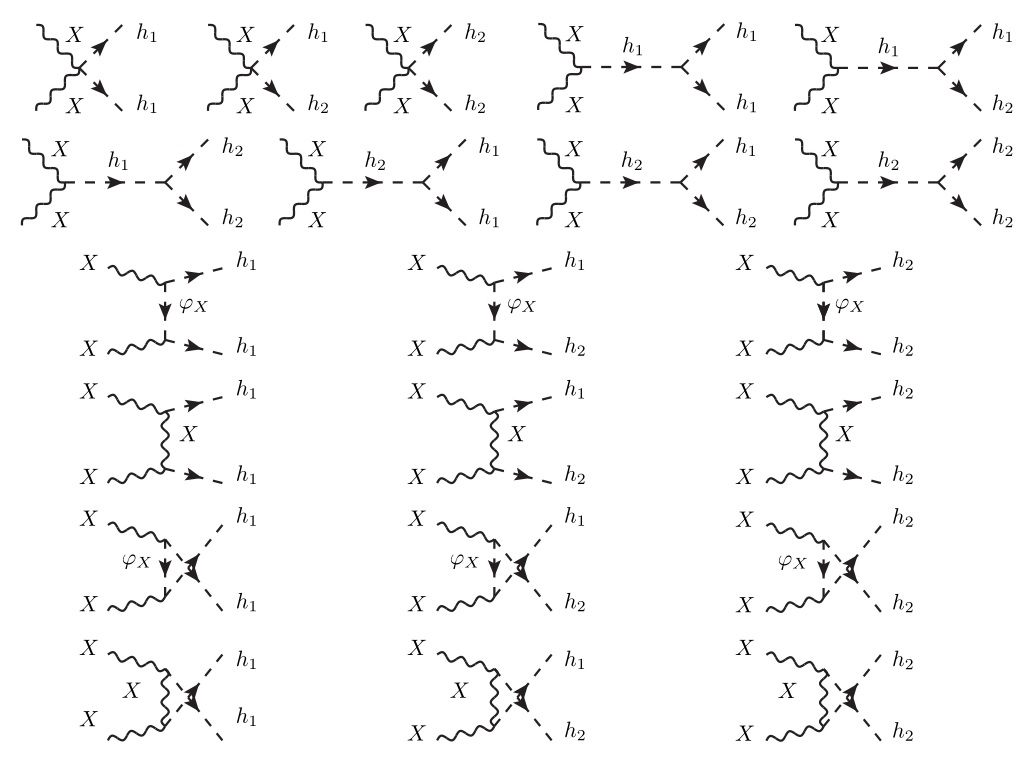}
\caption{Feynman diagrams for the annihilation process of the DM with scalars in the final state.}
\label{annichilazioni}
\end{figure}

\begin{figure}[tbh]
\centering
\includegraphics[scale=0.4]{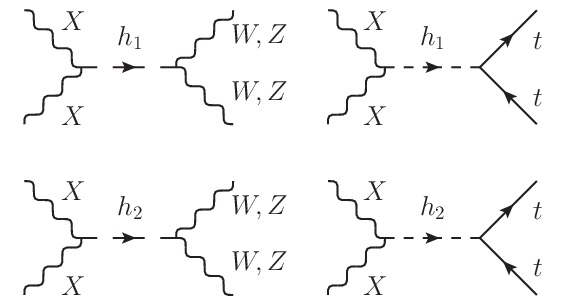}
\caption{Feynman diagrams for the annihilation process of the DM with W, Z or Top quark in the final state.}
\label{annichilazioni2}
\end{figure}

\begin{figure}[tbh]
\centering
\includegraphics[scale=0.4]{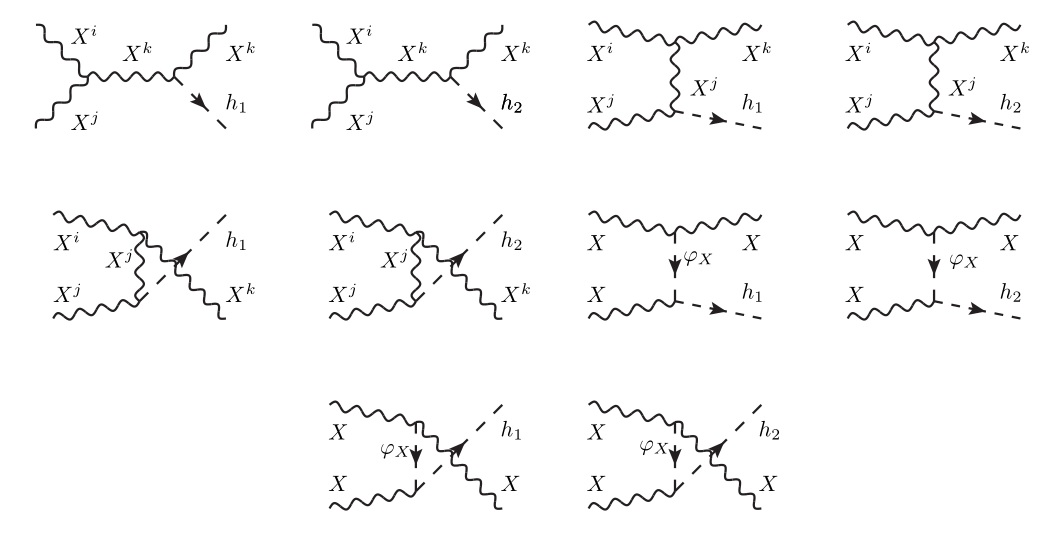}
\caption{Feynman diagrams for the semiannihilation process of the DM.}
\label{semiannichilazioni}
\end{figure}

We assume that DM is a thermal relic, so we are going to describe which processes are important during DM freeze-out. We have to consider the annihilation processes, i.e. $XX\rightarrow h_i h_j$ (Figure \ref{annichilazioni}), $XX\rightarrow WW$, $XX\rightarrow ZZ$ and $XX\rightarrow t\bar{t}$ (Figure \ref{annichilazioni2}). Moreover, we have to take in account the semiannihilation processes $XX\rightarrow X h_i$ (Figure \ref{semiannichilazioni}). Since DM is cold, we considered only the non-relativistic limit.

There are six contributions to the annihilation cross-section: the first one has two $h_1$ particles in the final state, the third one has two $h_2$ particles, while the second one has one of each scalar. The last three contributions are related respectively to the production of a couple of W, a couple of Z or a couple of Top quarks.

\begin{align}
\begin{split}
\sigma v_{\text{ann}}^{h_1,h_1}&=\frac{g_X^2}{110592 \pi  M_X^3} \sqrt{M_X^2-M_{h_2}^2}\\
&\times\left(\frac{16 g_X^2 \cos ^4\alpha \left(11 M_{h_2}^4-28 M_{h_2}^2 M_X^2+44 M_X^4\right)}{\left(M_{h_2}^2-2 M_X^2\right)^2}\right.\\
&+\frac{8 g_X M_X \cos ^2\alpha \left(M_{h_2}^2-10 M_X^2\right) \mathcal{A}_1 }{\left(4 M_X^2-M_{h_1}^2\right) \left(M_{h_2}^2-4 M_X^2\right) \left(M_{h_2}^2-2 M_X^2\right)}\\
&\left.+\frac{3 M_X^2 \mathcal{A}_1^2}{\left(M_{h_1}^2-4 M_X^2\right)^2 \left(M_{h_2}^2-4 M_X^2\right)^2}\right)
\end{split}
\end{align}

\begin{align}
\begin{split}
\mathcal{A}_1&=w (\lambda_{HS}-6\lambda_S) \left(3 M_{h_1}^2+M_{h_2}^2-16 M_X^2\right)\\
&+2 v \sin (2 \alpha ) \left(M_{h_1}^2 (6\lambda_H-3\lambda_{HS})-M_{h_2}^2 (6\lambda_H+\lambda_{HS})+16\lambda_{HS} M_X^2\right)\\
&-4 w \cos (2 \alpha ) \left(6\lambda_S M_{h_1}^2+\lambda_{HS} M_{h_2}^2-4 M_X^2 (\lambda_{HS}+6\lambda_S)\right)\\
&-3 v \sin (4 \alpha ) (2\lambda_H+\lambda_{HS}) \left(M_{h_1}^2-M_{h_2}^2\right)\\
&-3 w \cos (4 \alpha ) (\lambda_{HS}+2\lambda_S) \left(M_{h_1}^2-M_{h_2}^2\right)
\end{split}
\end{align}

\begin{align}
\begin{split}
&\sigma v_{\text{ann}}^{h_1,h_2}=\frac{g_X^2}{221184 \pi  M_X^8} \sqrt{M_{h_1}^4-2 M_{h_1}^2 \left(M_{h_2}^2+4 M_X^2\right)+\left(M_{h_2}^2-4 M_X^2\right)^2}\\
&\times\left(\frac{g_X^2 \sin ^2(2 \alpha )}{\left(M_{h_1}^2+M_{h_2}^2-4 M_X^2\right)^2}\right.\\
&\qquad \left(-224 M_X^6 \left(M_{h_1}^2+M_{h_2}^2\right)-12 M_X^2 \left(M_{h_1}^2-M_{h_2}^2\right)^2 \left(M_{h_1}^2+M_{h_2}^2\right)\right.\\
&\left.\qquad+\left(M_{h_1}^2-M_{h_2}^2\right)^4+M_X^4 \left(92 M_{h_1}^4-8 M_{h_1}^2 M_{h_2}^2+92 M_{h_2}^4\right)+704 M_X^8\right)\\
&+\frac{2 g_X M_X^3 \sin (2 \alpha ) \left(M_{h_1}^4+M_{h_2}^4-2 M_{h_1}^2 \left(M_{h_2}^2+M_X^2\right)-2 M_{h_2}^2 M_X^2+40 M_X^4\right) \mathcal{A}_2 }{\left(M_{h_1}^2-4 M_X^2\right) \left(4 M_X^2-M_{h_2}^2\right) \left(M_{h_1}^2+M_{h_2}^2-4 M_X^2\right)}\\
&+\left.\frac{3 M_X^6 \mathcal{A}_2^2}{\left(M_{h_1}^2-4 M_X^2\right)^2 \left(M_{h_2}^2-4 M_X^2\right)^2}\right)
\end{split}
\end{align}

\begin{align}
\begin{split}
\mathcal{A}_2&=v (6 \lambda_H-\lambda_{HS}) \left(M_{h_1}^2-M_{h_2}^2\right)\\
&-4 \lambda_{HS} v \cos (2 \alpha ) \left(M_{h_1}^2+M_{h_2}^2-8 M_X^2\right)\\
&+2 w \sin (2 \alpha ) (\lambda_{HS}+6 \lambda_S) \left(M_{h_1}^2+M_{h_2}^2-8 M_X^2\right)\\
&-3 v \cos (4 \alpha ) (2 \lambda_H+\lambda_{HS}) \left(M_{h_1}^2-M_{h_2}^2\right)\\
&+3 w \sin (4 \alpha ) (\lambda_{HS}+2 \lambda_S) \left(M_{h_1}^2-M_{h_2}^2\right)
\end{split}
\end{align}

\begin{align}
\begin{split}
\sigma v_{\text{ann}}^{h_2,h_2}&=\frac{g_X^2}{110592 \pi  M_X^3} \sqrt{M_X^2-M_{h_1}^2} \\
&\times\left(\frac{16 g_X^2 \sin ^4\alpha \left(11 M_{h_1}^4-28 M_{h_1}^2 M_X^2+44 M_X^4\right)}{\left(M_{h_1}^2-2 M_X^2\right)^2}\right.\\
&-\frac{8 g_X M_X \sin ^2\alpha \left(10 M_X^2-M_{h_1}^2\right)\mathcal{A}_3}{\left(M_{h_1}^2-4 M_X^2\right) \left(M_{h_1}^2-2 M_X^2\right) \left(4 M_X^2-M_{h_2}^2\right)}\\
&+\left.\frac{3 M_X^2 \mathcal{A}_3^2}{\left(M_{h_1}^2-4 M_X^2\right)^2 \left(M_{h_2}^2-4 M_X^2\right)^2}\right)
\end{split}
\end{align}

\begin{align}
\begin{split}
\mathcal{A}_3&=w (\lambda_{HS}-6 \lambda_S) \left(M_{h_1}^2+3 M_{h_2}^2-16 M_X^2\right)\\
&+2 v \sin (2 \alpha ) \left(M_{h_1}^2 (6 \lambda_H+\lambda_{HS})+M_{h_2}^2 (3 \lambda_{HS}-6 \lambda_H)-16 \lambda_{HS} M_X^2\right)\\
&+4 w \cos (2 \alpha ) \left(\lambda_{HS} M_{h_1}^2+6 \lambda_S M_{h_2}^2-4 M_X^2 (\lambda_{HS}+6 \lambda_S)\right)\\
&+3 v \sin (4 \alpha ) (2 \lambda_H+\lambda_{HS}) \left(M_{h_1}^2-M_{h_2}^2\right)\\
&+3 w \cos (4 \alpha ) (\lambda_{HS}+2 \lambda_S) \left(M_{h_1}^2-M_{h_2}^2\right)
\end{split}
\end{align}

\begin{equation}\sigma v_{\text{ann}}^{WW}=\frac{g_X^2 \sin ^2(2 \alpha ) \left(M_{h_1}^2-M_{h_2}^2\right)^2 \left(3 M_W^4-4 M_W^2 M_X^2+4 M_X^4\right) \sqrt{M_X^2-M_W^2}}{288 \pi  M_X v^2 \left(M_{h_1}^2-4 M_X^2\right)^2 \left(M_{h_2}^2-4 M_X^2\right)^2}\end{equation}

\begin{equation}\sigma v_{\text{ann}}^{ZZ}=\frac{g_X^2 \sin ^2(2 \alpha ) \left(M_{h_1}^2-M_{h_2}^2\right)^2 \left(3 M_Z^4-4 M_X^2 M_Z^2+4 M_X^4\right) \sqrt{M_X^2-M_Z^2}}{576 \pi  M_X v^2 \left(M_{h_1}^2-4 M_X^2\right)^2 \left(M_{h_2}^2-4 M_X^2\right)^2}\end{equation}

\begin{equation}\sigma v_{\text{ann}}^{TT}=\frac{g_X^2 \sin ^2(2 \alpha ) M_T^2 \left(M_{h_1}^2-M_{h_2}^2\right)^2 \left(M_T^2-M_X^2\right) \sqrt{M_X^2-M_T^2}}{288 \pi  M_X v^2 \left(M_{h_1}^2-4 M_X^2\right)^2 \left(M_{h_2}^2-4 M_X^2\right)^2}\end{equation}

The semiannihilation cross section is
\begin{align}
\begin{split}\sigma_\text{semiann}v&=\frac{g_X^4 \left(M_{h_1}^4-10 M_{h_1}^2 M_X^2+9 M_X^4\right)^{3/2} \sin^2 (\alpha )}{128 \pi  M_X^4 \left(M_{h_1}^2-3 M_X^2\right)^2}\\
&+\frac{g_X^4 \left(M_{h_2}^4-10 M_{h_2}^2 M_X^2+9 M_X^4\right)^{3/2} \cos^2 (\alpha )}{128 \pi  M_X^4 \left(M_{h_2}^2-3 M_X^2\right)^2}.
\end{split}
\end{align}

We observe that in the limit of small $\lambda_{HS}$, we get the same result of the approximated computation of \cite{stru}.

Calling $\sigma_\text{ann}$ and $\sigma_\text{semiann}$ the non-relativistic cross sections of these processes, we can say the experimental Dark Matter abundance is reproduced if~\cite{stru, planck}
\begin{equation}\sigma_\text{ann} v+\frac{1}{2}\sigma_\text{semiann}v = 2.2\times 10^{-26} \text{cm}^3/\text{s} =1.83 \times10^{-9} \text{GeV}^{-2},\end{equation}
where $v$ is the relative velocity between the initial particles.
We added a factor $1/2$ for the semi-annihilations because the number of DM particles drops only by one unit, so their contribution to the total annihilation of the DM is just one half of the contribution of the annihilations.
We averaged these cross sections over the polarizations of the vectors and over their SU(2)$_\text{X}$ indices.

\subsection{Corrections to the VEV of the Higgs}\label{HiggsVEV}
The VEV of the Higgs is fixed by the amplitude of the muon decay process.
The relation between the Higgs VEV and the Fermi constant $G_F$ is
\begin{equation}\frac{G_F}{\sqrt{2}}=\frac{1}{2v^2}(1+\Delta r),\end{equation}
where $\Delta r$ encloses all the contributions given by the corrections to the W boson propagator. In tree-level approximation we have $\Delta r=0$. The experimental value of the Fermi constant is $1.16637 \times 10^{-5} \text{GeV}^{-2}$, so, considering only tree-level diagrams we obtain $v\simeq 246.22\text{ GeV}$ from the previous relation.

We include one-loop corrections: $\Delta r$ is given by
\begin{equation}\Delta r^{\text{(1loop)}}=\Delta r^{\text{SM}}(M_h\rightarrow M_{h_1})\cos^2\alpha+\Delta r^{\text{SM}}(M_h\rightarrow M_{h_2})\sin^2\alpha.\end{equation}
where $\Delta r^{\text{SM}}$ is the known SM result \cite{sirlin} in the Landau gauge $\xi_W=\xi_Z=\xi_X=0$.

\section{Results}\label{results}
Now we have to write a system of equations, imposing that our observables agree with the experimental data. We introduced six parameters in this model, but the presently available observables that we computed give us only five conditions, so we choose $g_X$ as the only free parameter.
We find the values of $\lambda_H$, $\lambda_{HS}$, ${\mu^\ast}^2$, $v$ and $w$:
\begin{equation}
\begin{cases}
\frac{\partial V^{\text{1loop}}}{\partial h}=0\\
\frac{\partial V^{\text{1loop}}}{\partial s}=0\\
M_h^2=m_h^2+\Delta\Pi(p^2)=(125.6 \text{GeV})^2\\
\frac{1}{v^2\sqrt{2}}(1+\Delta r^{\text{(1loop)}})=G_F= 1.16637 \times 10^{-5} \text{GeV}^{-2}\\
\sigma_\text{ann} v+\frac{1}{2}\sigma_\text{semiann}v = 2.2\times 10^{-26} \text{cm}^3/\text{s} =1.83 \times10^{-9} \text{GeV}^{-2}
\end{cases}
\end{equation}

This system has to be solved in two cases, because we don't know which of the eigenvalues of the one-loop mass matrix corresponds to the Higgs. Then, we have to compute the solution for every value of $g_X$. We have a set of solutions showing us the values of the parameters of the model as function of $g_X$. With these data we plot the predicted cross section for the production of the new scalar as a function of the mass of the scalar itself, and a diagram of the spin-independent cross section for direct detection of the Dark Matter as a function of the mass of the DM particle and of the free parameter.

In the limit where we are taking in account only gauge interactions, our results reproduce those of \cite{stru}.

\begin{figure}[ht]
\centering
\includegraphics[height=0.38\textheight]{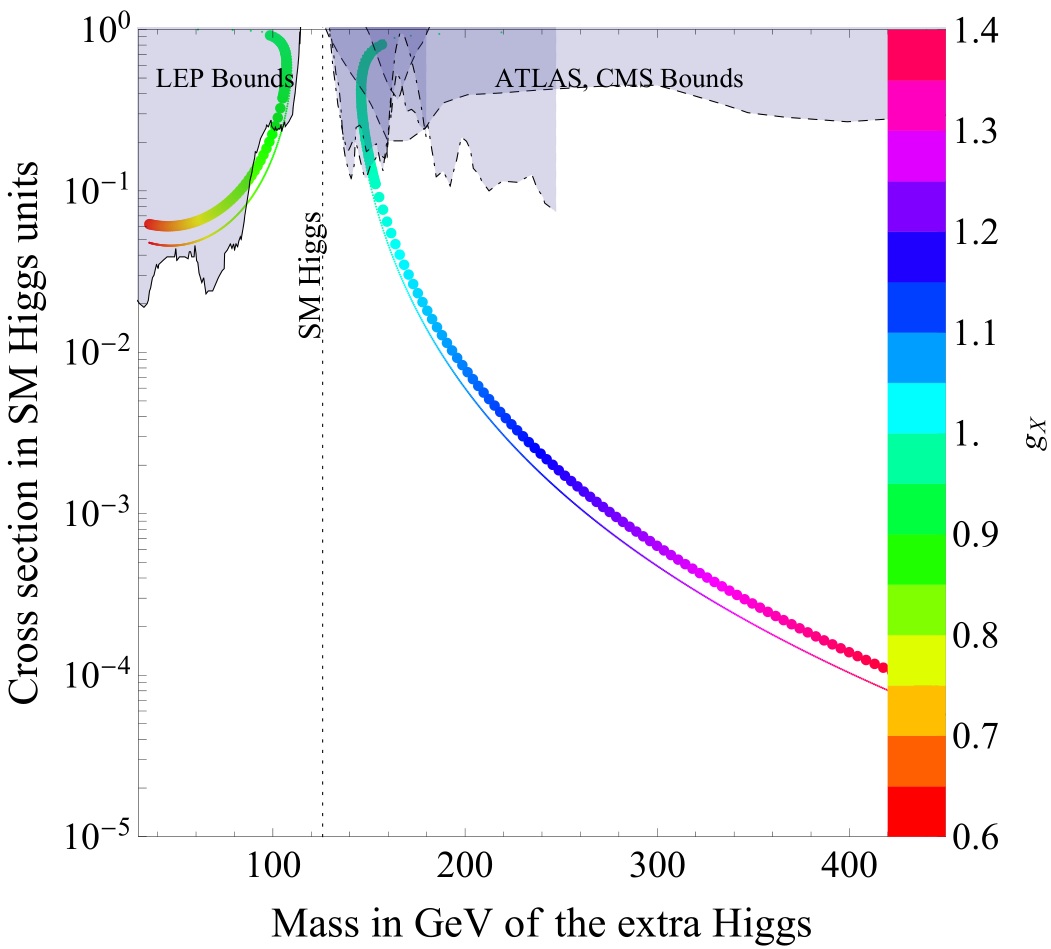}
\includegraphics[height=0.38\textheight]{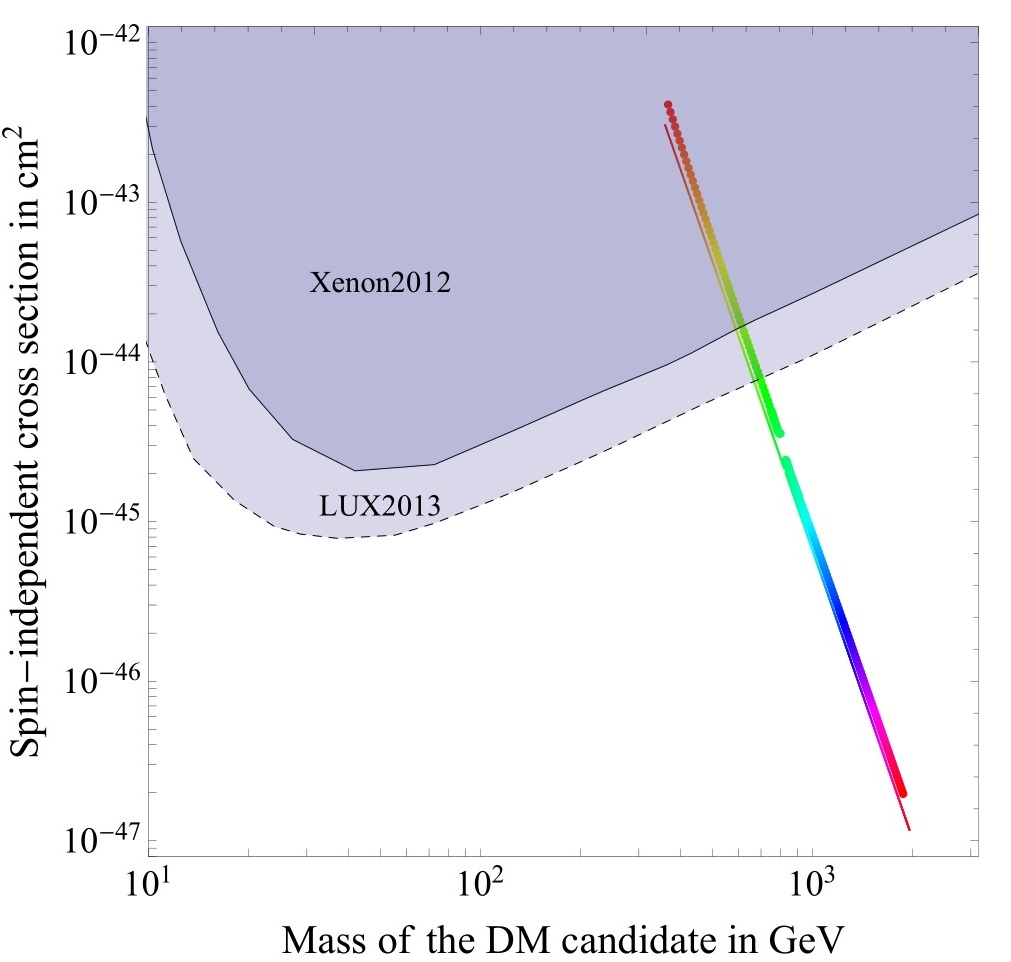}
\caption{Our final result: above, the prediction of the complete model for the production cross section of the new scalar. Below we report the prediction for the cross section for DM direct detection. These quantities are plotted as a function of the parameter $g_X$, that varies accordingly to the colors on the legend. For a comparison, in these diagrams we leave the data of the approximated case as smaller points. The grey areas are excluded by LEP or CMS and ATLAS experiments for the diagram above, while the bounds comes from XENON2012 and from LUX2013 experiments for the diagram below (see Section \ref{results} for more detail on the bounds).}
\label{completi}
\end{figure}

The plots report the bounds given by LEP experiments for energies lower than the Higgs mass and by ATLAS and CMS experiments for greater energies. The bounds of the LEP experiments are the 95\% confidence level of the ratio $\sigma/\sigma^{\text{SM}}$, where $\sigma$ is the measured production cross section of the Higgs, while $\sigma^{\text{SM}}$ is its theoretical value for the SM \cite{bounds1}. The bounds given by ATLAS and CMS at large mass are the 95\% CL upper limits on the production cross section of a Higgs boson: $h\rightarrow WW$ searches are plotted as dashed curves and $h\rightarrow WW$ searches as dot-dashed curves \cite{atlas, bounds2, bounds4}.
Below, in Figure \ref{completi}, we present the spin independent cross section for direct detection in the complete case. The bounds are the 90\% confidence level of spin-independent DM-nucleon scattering cross section. The model is not excluded by XENON2012 and LUX2013 data for $g_X\gtrsim0.8$ \cite{xenon, lux}.

\section{RGE analysis}
After finding the values of the parameters of the model around the weak scale, now we have to explore the behaviour of these parameters at high energy. To do this, first we must find the renormalization group equations (RGEs) for the theory. We can get these equation starting from those of the Standard Model. In particular, the scalar quartic coupling has an extra term due to the interaction between the Higgs boson and the new scalar boson:
\begin{equation}
(4\pi)^2\frac{d\lambda_H}{d \ln\mu}=(12 g_t^2-\frac{9}{5}g_1^2-9 g_2^2)\lambda_H-6g_t^4+\frac{27}{200}g_1^4+\frac{9}{20}g_2^2g_1^2+\frac{9}{8}g_2^4+24\lambda_H^2+2\lambda_{HS}^2
\end{equation}
The new coupling constants RGEs are
\begin{equation}
(4\pi)^2\frac{dg_X}{d \ln\mu}=-\frac{43}{6}g_X^3-\frac{1}{(4\pi)^2}\frac{259}{6}g_X^5
\end{equation}
\begin{equation}
(4\pi)^2\frac{d\lambda_{HS}}{d \ln\mu}=\lambda_{HS}(6g_t^2-\frac{9}{2}g_X^2-\frac{9}{10}g_1^2-\frac{9}{2} g_2^2+12\lambda_H+12\lambda_S)-4\lambda_{HS}^2
\end{equation}
\begin{equation}
(4\pi)^2\frac{d\lambda_S}{d \ln\mu}=-9g_X^2\lambda_S+\frac{9}{8}g_X^4+2\lambda_{HS}^2+24\lambda_S^2
\end{equation}
The other RGEs of the SM are not affected by the new sector of this model, so they remain exactly the same. In our computation we set the mass the Higgs boson equal to 125.9 GeV and the Top quark mass equal to 173 GeV.

\begin{figure}[ht]
\centering
\includegraphics[height=0.38\textheight]{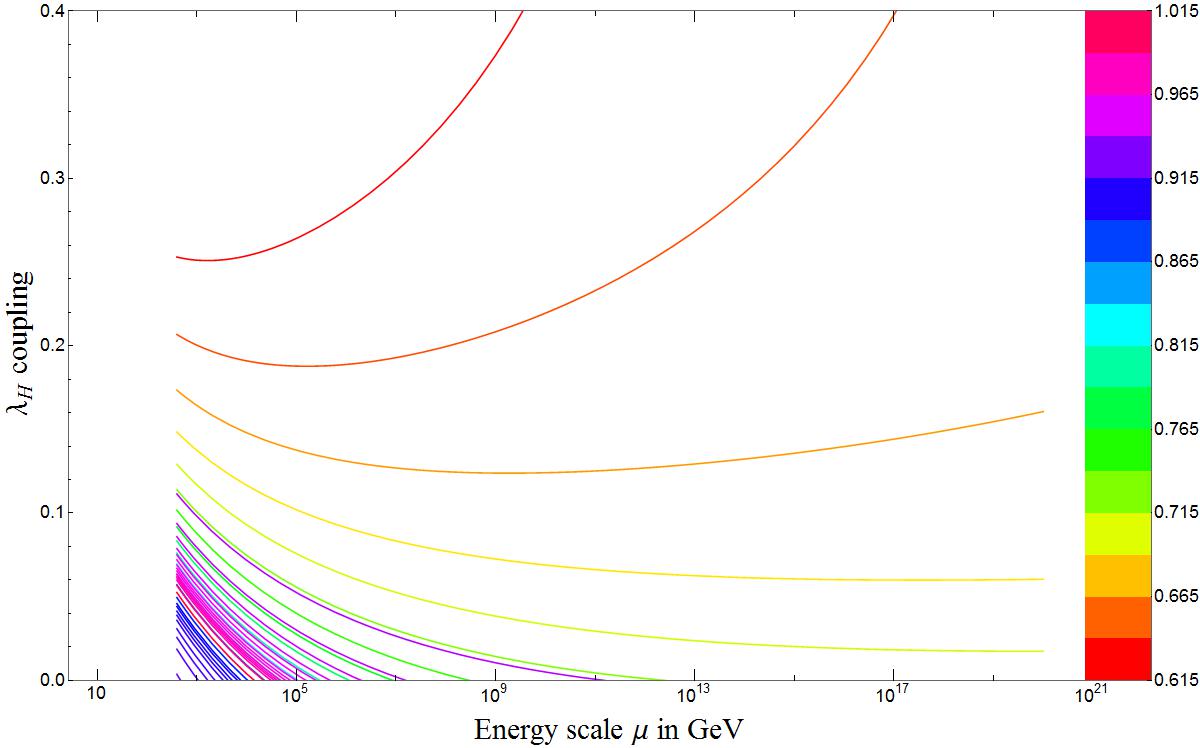}
\caption{Result of the RGE analysis for different values of the free parameter $g_X$: the running quartic coupling of this model remains positive at all the energy scales when $g_X\lesssim 0.695$. For bigger values of this parameter, that correspond to a heavier new scalar boson, there is an energy scale for which the quartic coupling $\lambda_H$ becomes negative, causing the instability of the vacuum. }
\label{RGE2}
\end{figure}

The results of are represented in Figure \ref {RGE2}. First, we considered the case in which $g_X=1$ at the weak scale. We observe that for this value the problem of the vacuum instability of the SM is not solved yet. Since in this article we considered $g_X$ as a free parameter of the theory, we studied if there is an interval for which the quartic coupling of the Higgs boson never becomes negative, for every energy scale. In the second diagram we see the running of the quartic scalar coupling from the weak scale to the Planck scale for a restricted range of values for the parameter $g_X$. We can say that the minimum is stable for $g_X\lesssim 0.695$ that corresponds to a mass of the new scalar lower than 53 GeV.

\section{Conclusions}
We considered an extension of the SM that describes the Dark Matter and proposes a reinterpretation to the hierarchy problem.

In the context of ``finite naturalness'', we introduced a model without a mass term for the Higgs. The masses of the particles arise from a Coleman-Weinberg mechanism, so spontaneous symmetry breaking does not occur at tree-level, but is generated by the radiative corrections to the theory. We supposed that there is a new particle $S$, scalar doublet under an extra group SU(2)$_{X}$, and new vector bosons $X$ of the same gauge group. The only communication between this new sector and the SM is through the so-called ``Higgs portal'', that is the quartic vertex between two Higgs fields and two $S$ fields.
The VEVs of the two scalars are fixed by the one-loop potential; the interactions with the scalars give mass to all the particles of the model.

We introduced the vector boson $X$ of SU(2)$_{X}$, that is a good candidate to represent the Dark Matter. It has a mass of about 1 TeV, and if we make a rough estimate, this is the order of magnitude of the scale where the mass of the DM particle is expected, assuming it is a thermal relict. Furthermore, this particle has to be stable. Some theories have to introduce special symmetries with the specific purpose of keeping the DM particle stable. In our model, $X$ vectors are automatically stable, because of the gauge symmetry and because of the particle content of the theory. 

Another peculiarity of this simple model is the presence of only one free parameter. The other parameters introduced in the model are fixed by the experimental values of the DM cosmological abundance, of the Fermi constant and of the Higgs mass.

The original work presented in this article consisted in performing for the first time a precise computation of the predictions of the model for the LHC and for direct detection experiments. The new computation includes for the first time a full one-loop computation of the scalar masses and of the effective potential, and a full tree-level computation of the DM annihilations and semi-annihilations relevant for the thermal DM abundance.

In our computation the numerical solving algorithm of the system converges to an acceptable solution for a bigger number of values of the free parameter than the previous works on this model. For example, in the diagram of the direct dectection cross section of the DM of \cite{stru}, there were no points for the values $g_X\approx 0.9$, that corresponds to a mass of the new scalar almost equal to the Higgs mass.

Given the mass $M_s$ of the extra scalar, the cross section for its production for LHC increases by a factor $\approx$1.3 with respect to the approximated computation. This cross section is compatible with the experiments in a small range around $g_X\approx0.9$ when the new scalar is lighter than Higgs, and for $g_X\gtrsim 1.0$ when it is heavier.

From the LUX2013 and XENON2012 we get a constrain on the free parameter: the prediction for the DM direct detection is compatible with the bounds for $g_X\gtrsim 0.8$, that corresponds to a mass of the DM particle $\gtrsim 700$ GeV.

We considered also the behaviour of the coupling constant at high energy, studying their running through the RGEs. We can see that there are some values of the free parameter, that is the gauge coupling $g_X$, the Higgs quartic coupling stays automatically positive at all the energy scales: this case occurs for $g_X\lesssim0.695$, that is for a new scalar lighter than 53 GeV. Obviously this case has already been disfavoured by the other analysis done in this article, so the question about the instability of the vacuum remains open, and more new physics is needed to solve this problem.

The collaborations of the new experiments LZ and XENON1T claim that their detectors will probe spin indipendent cross sections of the DM direct detection almost 2 or 3 orders of magnitude lower than the actual values reached by LUX and XENON100. This sensitivity will surely be enough to directly detect the new DM particle of this model in a mass range around 1 TeV.
Considering the substantial increase in luminosity of LHC and simulations made by ATLAS and CMS (for example see \cite{atlas2}), we expect that the new run of the LHC will see the new scalar boson if it has a mass $\lesssim 200$ GeV.

\section*{Acknowledgments}
We thank A. Strumia for the collaboration and for the useful discussions.


\begin{thebibliography}{100}
\bibitem{atlas} ATLAS Collaboration, \emph{Observation of a new particle in the search for the Standard Model Higgs boson with the ATLAS detector at the LHC}, Phys. Lett. B 716 (2012) 1 [arXiv:1207.7214].
\bibitem{cms} CMS Collaboration, \emph{Observation of a new boson at a mass of 125 GeV with the CMS experiment at the LHC}, Phys. Lett. B 716 (2012) 30 [arXiv:1207.7235].
\bibitem{thoft} G. 't Hooft, \emph{Naturalness, Chiral Symmetry and Spontaneous Chiral Symmetry Breaking} in \emph{Recent Developments in Gauge Theories}, ISBN 978-0-306-40479-5.
\bibitem{fari}  M. Farina,  D. Pappadopulo and  A. Strumia, \emph{A modified naturalness principle and its experimental tests}, [arXiv:1303.7244]
\bibitem{coleman} S.R. Coleman and E.J. Weinberg, Phys.\ Rev.\ D 7 (1973) 1888. E. Gildener and  S. Weinberg, Phys.Rev.D 13 (1976) 3333.
\bibitem{tipocoleman1} R. Hempfling, \emph{The Next-to-Minimal Coleman-Weinberg Model}, Phys. Lett. B 379 (1996) 153 [hep-ph/9604278].
\bibitem{tipocoleman2} J. P. Fatelo, J. M. Gerard, T. Hambye and J. Weyers, \emph{Symmetry Breaking Induced by Top Quark Loops} Phys. Rev. Lett. 74 (1995) 492.
\bibitem{tipocoleman3} T. Hambye, Phys. Lett. B 371 (1996) 87, \emph{Symmetry breaking induced by top quark loops from a model without scalar mass}, [hep-ph/9510266].
\bibitem{tipocoleman4} W. F. Chang, J. N. Ng and J. M. S. Wu, Phys. Rev. D 75 (2007) 115016, \emph{Shadow Higgs from a scale-invariant hidden U(1)$_s$ model} [hep-ph/0701254].
\bibitem{tipocoleman5} R. Foot, A. Kobakhidze and R. R. Volkas, Phys. Lett. B 655 (2007) 156 [arXiv:0704.1165].
\bibitem{tipocoleman6} R. Foot, A. Kobakhidze, K. L. McDonald, R. R. Volkas, \emph{Electroweak Higgs as a pseudo-Goldstone boson of broken scale invariance}, Phys. Rev. D 77 (2008) 035006 [arXiv:0709.2750].
\bibitem{tipocoleman7} S. Iso, N. Okada, Y. Orikasa, \emph{The minimal B-L model naturally realized at TeV scale}, Phys. Rev. D 80 (2009) 115007 [arXiv:0909.0128].
\bibitem{tipocoleman8} S. Iso and Y. Orikasa, \emph{TeV Scale B-L model with a flat Higgs potential at the Planck scale -- in view of the hierarchy problem} PTEP 2013 (2013) 023B08 [arXiv:1210.2848].
\bibitem{tipocoleman9} C. Englert, J. Jaeckel, V. V. Khoze and M. Spannowsky, \emph{Emergence of the Electroweak Scale through the Higgs Portal}, [arXiv:1301.4224].
\bibitem{tipocoleman10}E. J. Chun, S. Jung, and H. M. Lee, \emph{Radiative generation of the Higgs potential}, Phys. Lett. B 725 (2013) 158 [arXiv:1304.5815].
\bibitem{tipocoleman12}  M. Cirelli, N. Fornengo and A. Strumia, \emph{Minimal Dark Matter}, Nucl. Phys. B 753 (2006) 178 [hep-ph/0512090].
\bibitem{estonia1} M. Heikinheimo, A. Racioppi, M. Raidal, C. Spethmann and K. Tuominen, \emph{Physical Naturalness and Dynamical Breaking of Classical Scale Invariance}, [arXiv:1304.7006].
\bibitem{estonia2} M. Heikinheimo, A. Racioppi, M. Raidal, C. Spethmann and K. Tuominen, \emph{Dark Supersymmetry}, [arXiv:1305.4182.]
\bibitem{estonia3} E. Gabrielli, M. Heikinheimo, K. Kannike, A. Racioppi, M. Raidal, and C. Spethmann, \emph{Towards Completing the Standard Model: Vacuum Stability, EWSB and Dark Matter}, [arXiv:1309.6632]
\bibitem{hamb} T. Hambye, \emph{Hidden vector Dark Matter}, [arXiv:0811.0172].
\bibitem{carone}  C.~D.~Carone and R.~Ramos, \emph{Classical scale-invariance, the electroweak scale and vector dark matter},  Phys.\ Rev.\ D 88, n.\ 5, 055020 (2013) [arXiv:1307.8428].
\bibitem{stru} T. Hambye and A. Strumia, \emph{Dynamical generation of the weak and Dark Matter scale}, [arXiv:1306.2329].
\bibitem{planck}  Planck collaboration, \emph{Planck 2013 results. XVI. Cosmological parameters}. [arXiv:1303.5076].
\bibitem{sirlin}A. Sirlin, R. Zucchini, \emph{Dependence of the Higgs coupling} $\bar{h}_{\overline{\text{MS}}}$ on $m_H$ and the possible onset of new physics, Nuclear Physics B 266,  389
\bibitem{bounds1} ALEPH, DELPHI, L3 and OPAL Collaborations, The LEP Working Group for Higgs Boson Searches, \emph{Search for the Standard Model Higgs Boson at LEP}, Physics Letters B
565 (2003), 61 [arXiv:hep-ex/0306033]
\bibitem{bounds2}  CMS Collaboration, \emph{Observation of a new boson with mass near 125 GeV in pp collisions at sqrt(s) = 7 and 8 TeV}, JHEP 06 (2013) 081, [arXiv:1303.4571].
\bibitem{bounds4}M. Flechl, on behalf of the ATLAS, CMS collaborations, \emph{BSM Higgs results from ATLAS and CMS}, [arXiv:1307.4589]
\bibitem{xenon} XENON100 Collaboration, \emph{Dark Matter Results from 225 Live Days of XENON100 Data}, Phys. Rev. Lett. 109, 181301 (2012), [arXiv:1207.5988]
\bibitem{lux} LUX Collaboration, \emph{First results from the LUX dark matter experiment at the Stanford Underground Research Facility}, Phys. Rev. Lett. 112, 091303 (2014) [arXiv:1310.8214]
\bibitem{atlas2} ATLAS Collaboration, \emph{Beyond-the-Standard-Model Higgs boson searches at a High-Luminosity LHC with ATLAS}, ATLAS-PHYS-PUB-2013-016
\end{thebibliography}
\end{document}